\documentclass[iop]{emulateapj}

\usepackage{textcomp}
\usepackage{array,multirow}
\usepackage{natbib}
\usepackage{amsmath}
\usepackage{xcolor}

\defcitealias{KH09}{KH09}
\defcitealias{Mut13}{M13}
\defcitealias{Gab13}{MG3}
\defcitealias{Air15}{A15}
\defcitealias{Air12}{A12}

\begin{document}

\title{Evolution of Black Hole and Galaxy Growth in a Semi-numerical Galaxy Formation Model}

\author{Mackenzie L. Jones\altaffilmark{1,2}, Ryan C. Hickox\altaffilmark{2}, Simon J. Mutch\altaffilmark{3}, Darren J. Croton\altaffilmark{4}, Andrew F. Ptak\altaffilmark{5}, Michael A. DiPompeo\altaffilmark{2}}
\affil{$^{1}$Harvard-Smithsonian Center for Astrophysics, 60 Garden Street, Cambridge, MA 02138, USA}
\affil{$^{2}$Department of Physics and Astronomy, Dartmouth College, Hanover, NH 03755, USA}
\affil{$^{3}$School of Physics, University of Melbourne, Parkville, Victoria 3010, Australia}
\affil{$^{4}$Centre for Astrophysics \& Supercomputing, Swinburne University of Technology, PO Box 218, Hawthorn, VIC 3122, Australia}
\affil{$^{5}$NASA Goddard Space Flight Center, Code 662, Greenbelt, MD 20771, USA}

\begin{abstract}

We present a simple semi-numerical model designed to explore black hole growth and galaxy evolution. This method builds on a previous model for black hole accretion that uses a semi-numerical galaxy formation model and universal Eddington ratio distribution to describe the full AGN population by independently connecting galaxy and AGN growth to the evolution of the host dark matter halos.
We fit observed X-ray luminosity functions up to a redshift of $z \sim 4$, as well as investigate the evolution of the Eddington ratio distributions. We find that the Eddington ratio distribution evolves with redshift such that the slope of the low-Eddington accretion rate distribution increases with cosmic time, consistent with the behavior predicted in hydrodynamical simulations for galaxies with different gas fractions. We also find that the evolution of our average Eddington ratio is correlated with observed star formation histories, supporting a picture in which black holes and galaxies evolve together in a global sense. We further confirm the impact of luminosity limits on observed galaxy and halo properties by applying selection criteria to our fiducial model and comparing to surveys across a wide range of redshifts.

\end{abstract}

\keywords{galaxies: active}

\section{Introduction}\label{sec:intro}

Despite significant progress in establishing a generalized model of the formation and evolution of galaxies with respect to their host dark matter halos, it is still uncertain how the growth of supermassive black holes (SMBH) fit into this evolutionary scheme and what impact actively accreting supermassive black holes (active galactic nuclei; AGN) have on their hosts (for reviews, see e.g., \citealt{Sil12, Ale12,Fab12}). 

There is observational evidence of a co-evolution between AGN and their host galaxies, as well as theoretical models that explore the impact of feedback (both stellar and AGN) on galaxy growth (e.g., \citealt{Bow06,Cro06,Vol15}). \citet{Hec04} find that the volume-averaged galaxy-black hole growth rate, for a range of black hole masses, is consistent with the observed black hole-spheroid mass relationship. Additional evidence shows that black hole mass is correlated with stellar bulge properties (e.g., \citealt{Kor13}). There is also a similarity in the growth histories of black hole accretion and star formation across cosmic time, including where they peak at $z \sim 1-2$ (e.g., \citealt{Air10,Raf11,Mul12}). \citet{Che13} observed that the average black hole accretion rate is directly proportional to the star formation rate in star forming galaxies, although this may be a secondary effect due to the dependence of black hole accretion on stellar mass (\citealt{Yan17}). However, other than a common supply of cold gas on kpc scales, the physical processes linking black hole growth to galaxy evolution are still not well understood (e.g., \citealt{Dim05,Hop06,Ale12}).

Disentangling this co-evolution is made more difficult by the presence of observational biases (e.g., selection effects, obscuration from gas and dust, dilution from host galaxy emission) which may obscure these relationships. 
How an AGN is selected in a sample can be strongly impacted by these observational biases (see \citealt{Pad17} for a review; e.g., \citealt{Lau07}). Some selection criteria are more complex than others; for example, color selection and emission line ratio selection rely on detailed decompositions to separate the AGN from the host galaxy properties. Others are more straightforward, such as X-ray AGN selection based on an observed or absorption corrected luminosity limit.

In principle it is possible to correct for the presence of biases in most observations by assuming an underlying black hole accretion distribution and forward modeling the observations. Using SDSS star forming galaxies, \citet{Jon16} uncovered the underlying Eddington ratio distribution (distribution of the ratio of the instantaneous luminosity to the maximum possible accretion given by the Eddington limit; ${L}/{L}_\text{Edd}$), using a straight-forward model for black hole accretion to correct for the effects of dilution that are more prevalent in galaxies with star formation. This simple model for black hole accretion was also tested in \citet{Jon17} at $z = 0$ to place AGN in simulated galaxies from the \citet{Mut13} semi-numerical galaxy formation model. The results of this work showed that selection criteria introduce potential bias by selecting different host galaxy and dark matter halo properties (see also e.g., \citealt{Aza15,Air15}). The Eddington ratio distribution is observed directly in the X-rays to be consistent with a power law with a potential cutoff at high ${L}/{L}_\text{Edd}$ (e.g., \citealt{Air10,Air15}). Many theoretical models assume a simple functional form of the Eddington ratio distribution (e.g., \citealt{Con13,Hop09,Hic14,Jon16}), however, it is uncertain whether it should be dependent on mass and/or vary across cosmic time (e.g., \citealt{Cap15,Wei17,Ber18}).

In this work, we use an updated version of the \citet{Jon17} model for black hole accretion to investigate the black hole-galaxy connection as well as the role AGN selection criteria play in the observed properties of galaxy and dark matter halos. This method is computationally less expensive than in \citet{Jon17} and includes evolution of the AGN, galaxy, and dark matter halo properties up to a redshift of $z \sim 4$. We discuss how the simulation is constructed in Section \ref{sec:method}, including the modeling of the evolution of the observed X-ray luminosity functions (Section \ref{ssec:luminosity}). Our results are found in Section \ref{sec:xray}, where we also discuss potential evidence for AGN-galaxy co-evolution (Section \ref{ssec:evol}) and the change in galaxy properties observed when imposing selection criteria. Our conclusion and a summary of our results are provided in Section \ref{sec:con}. Throughout this paper, we assume a 1-year Wilkinson Microwave Anisotropy Probe (WMAP1; \citealt{Spe03}) cold dark matter (CDM) cosmology with $\Omega_m$ = 0.25, $\Omega_\lambda$ = 0.75, and $\Omega_b$ = 0.045. All results are shown with a Hubble constant of $\text{h} = 0.7$, where $\text{h} \equiv \text{H}_0/100$ km s$^{-1}$ Mpc$^{-1}$, and all luminosities are presented in erg s$^{-1}$.

\section{Simulation Methodology}\label{sec:method}

\subsection{Dark Matter and Galaxies}\label{ssec:gal}

As in \citet{Jon17}, we use a semi-numerical galaxy formation model (\citealt{Mut13}) as the foundation of our simulation. This model connects galactic growth to the formation history of the N-body Dark Matter Millennium Simulation (\citealt{Spr05}) with a prescribed baryonic growth function and a physics function. These simple analytic functions dictate the availability of baryonic material to be used by stars and the efficiency by which this process occurs. The merits of using this analytic solution, as well as a description of how we repopulate the output star formation rates, are described in \citet{Jon17}. 

We use 41 of the 63 available Millennium snapshots corresponding to a redshift range of $0\le z<4.12$ in steps of $\sim$ 200 -- 350 Myr (the remaining 22 snapshots are at redshifts $> 4.12$). This expands on the work in \citet{Jon17}, in which we limited our investigation to $z = 0$. Within this redshift range, the galaxy formation model is able to accurately reproduce the stellar mass function and its evolution to $z \sim 4$. These evolving simulated stellar mass functions inform our black hole mass distributions and so are used as a constraint in determining the best black hole accretion distribution model at each redshift, as discussed in Section \ref{ssec:agn}. For more information on the galaxy formation model, see \citet{Mut13}. 

\subsection{Adding in AGN activity}\label{ssec:agn}

Our simulated galaxy sample consists of approximately 24 million galaxies per snapshot. We add an AGN counterpart to each galaxy using the following method.
We first calculate black hole masses using the \citet{Har04} black hole-bulge relationship using the total galaxy stellar mass (M*) from the \citet{Mut13} galaxy formation model as a proxy for the bulge mass (\citealt{Kor13}). The merits of using this relationship to calculate our black hole masses are discussed in \citet{Jon17} and summarized below.

\begin{figure}[!t]
\begin{center}
\resizebox{85mm}{!}{\includegraphics{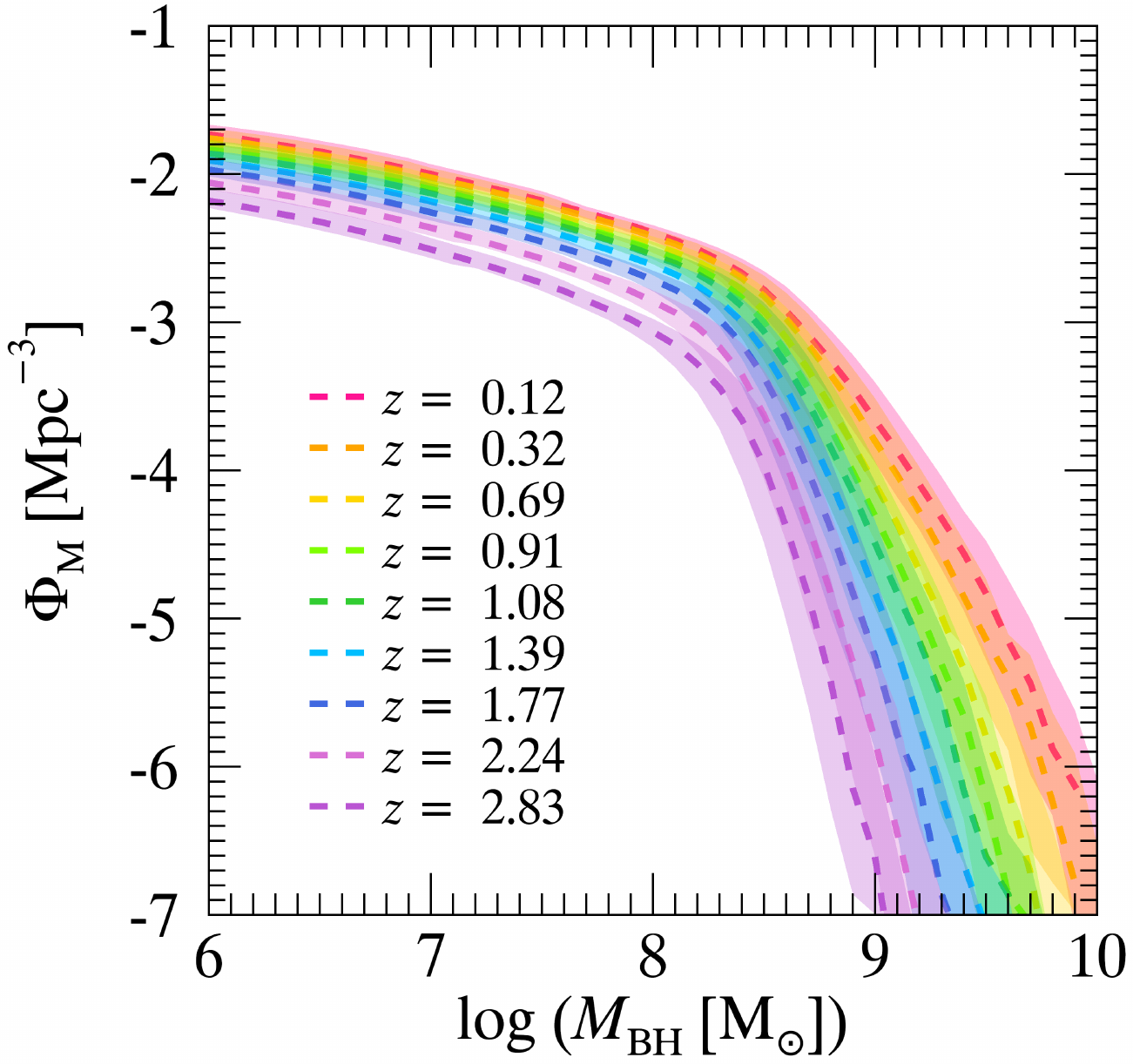}} \\
\caption{Evolution of the black hole mass function calculated directly from the \citet{Mut13} evolving galaxy stellar mass functions using the \citet{Har04} black hole-bulge relationship (dashed lines) with galaxy stellar mass used as a proxy for bulge mass (\citealt{Kor13}). $1\sigma$ error associated with the observed scatter of the black hole-mass relationship is shown in corresponding color bands. \label{fig:BHM}}
\end{center}
\end{figure}

The significant uncertainty and scatter found in measuring the relationship between black hole mass and total stellar mass has led to disagreements on the necessity of adding in an evolutionary component (e.g., \citealt{Dec10,Mer10,Scz11,Ben11,Woo13,Deg15,Sha16}). It is possible, however, that the differences observed for redshift evolution in this relationship may be caused by sample selection bias, rather than an intrinsic evolution. The observed differences in the black hole mass-stellar mass relationship based on galaxy type adds a further complication (e.g., \citealt{Rei15,Bal16}). Since the \citet{Har04} relationship bisects those found for ellipticals and AGN, we adopt this non-evolving model not only for simplicity, but as a way to handle the large intrinsic scatter in the relationship and limit the impact of potential observational biases. We investigate the impact of scatter observed in the relationship between black hole mass and stellar mass and the error associated with \citet{Har04} by adding a scatter of $\sim0.3$ dex to our black hole masses and running a Monte Carlo simulation with an additional uncertainty corresponding to the \citet{Har04} published errors to demonstrate the variance of our black hole mass functions (Figure \ref{fig:BHM}).
As a further check, we test an alternate black hole-stellar mass relationship with moderate evolution (\citealt{Mer10}). We evaluate the merit of adding a black hole evolution by fitting to the X-ray luminosity function as we do for the non-evolving case (described in detail below). However, we do not find a significant enough improvement to justify this additional complexity. We thus select the simplest form of the black hole-stellar mass relationship to limit the number of free parameters and avoid adding any additional degeneracies.

Building on the method in \citet{Jon17}, we determine the black hole accretion by convolving a broad Eddington ratio distribution with the simulated black hole mass functions to reproduce a bolometric luminosity function. We first fit the black hole mass function calculated from the \citet{Mut13} galaxies using a Schechter function (power law with exponential cutoff). While Eddington ratio distributions have been defined using a variety of functional forms (e.g., \citealt{Hop09Her,Nov11,Con13,Hic14,Jon17}), our Eddington ratio distributions are defined by a double power law in order to reproduce the evolution of the \citet{Air15} X-ray luminosity functions:
\begin{equation}\label{equ:bpl}
\frac{dt}{d\log \text{L}_{\textrm{bol}}}= \phi^*\left[\left(\frac{\text{L}_{\textrm{bol}}}{\lambda_{\textrm{break}}\text{L}_{\textrm{Edd}}}\right)^{-\alpha_1}+\left(\frac{\text{L}_{\textrm{bol}}}{\lambda_{\textrm{break}}\text{L}_{\textrm{Edd}}}\right)^{-\alpha_2}\right],
\end{equation}
where ${L}_{\text{bol}}$ is the bolometric luminosity, ${L}_{\text{bol}}/{L}_{\text{Edd}}$ is the Eddington ratio, $\phi^*$ is the amplitude, $\lambda_{\text{break}}$ marks the position of the break in the distribution, and $\alpha_1$, $\alpha_2$ are the power law slopes. In this work, our amplitude is set by the minimum Eddington ratio, $\lambda_\text{min}$, which is chosen such that the integral of our Eddington ratio distribution is one. This method is computationally inexpensive, so we can generate a wide range of Eddington ratio distributions to be used in the convolution in order to better reproduce observed luminosity functions.


\begin{figure*}[!t]
\begin{center}
\resizebox{150mm}{!}{\includegraphics{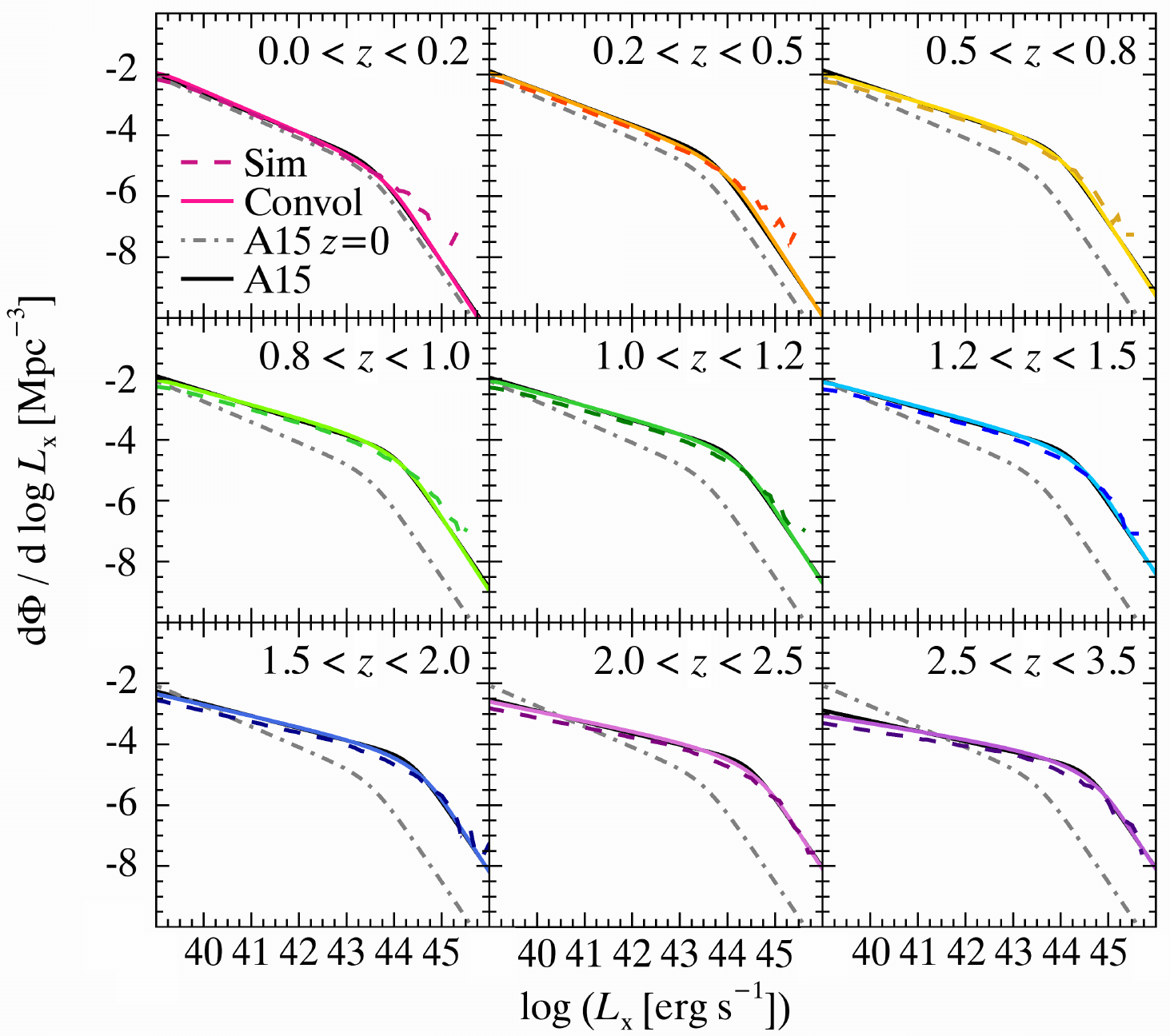}} \\
\caption{ The evolution of the X-ray luminosity function (XLF) from this work compared to the parametrization from \citet{Air15} for a selection of nine redshifts. We fit the \citet{Air15} XLF (solid black) at each redshift by convolving our simulated black hole mass function with an intrinsic Eddington ratio distribution that is defined by a broken power law (solid light colors). We select the best-fit Eddington ratio distribution parameters from the convolution using a least squares fit and build a simulated sample of AGN (dash dark corresponding colors) based on this black hole accretion activity. The \citet{Air15} $z = 0$ line (dot dash grey) is included for clarity. We find good agreement between the convolution and the simulation to the \citet{Air15} XLFs. Deviation of the simulated XLF at higher luminosities for lower redshifts is due to low number statistics, which is not surprising since very active AGN are much rarer in the local universe. \label{fig:XLF}}
\end{center}
\end{figure*}

\subsection{Recovering the AGN X-ray Luminosity Function}\label{ssec:luminosity}

In order to determine the intrinsic black hole activity, we can compare the output of our convolution to observed luminosity functions, such as the AGN X-ray luminosity functions (XLFs) of \citet{Air15}. We convert our bolometric luminosities into hard ($2-10$ keV) X-ray luminosities using a constant bolometric correction of ${k}_\text{bol}=44$ \citep{Don18}, although we did consider a variety of corrections, both constant as well as tied to luminosity (e.g., \citealt{Mar04,Hop07,Run12}). This differs from the prescription outlined in \citet{Jon17}, in that here we do not assume a connection between the bolometric corrections and the intrinsic Eddington ratio, as observed in the work by \citet{Lus12}. We assume a constant correction for simplicity because we find that using an Eddington ratio dependent bolometric correction requires additional free parameters that make our model unnecessarily complicated in order to fit the \citet{Air15} XLFs. We test our simulation with an accretion-dependent bolometric correction (\citealt{Lus12}), and find that the results of our analysis are consistent with a constant bolometric correction, although prone to more degeneracies in this instance.
Further simplifying our process, we no longer require the addition of obscuration (outlined in \citealt{Jon17}) to compare to the observed XLFs, as obscuration is corrected for in the \citet{Air15} XLFs. 

We fit the \citet{Air15} AGN XLF parametrization at each of our redshift snapshots by running a least squares fit with the convolved Eddington ratio distribution and black hole mass function. To determine the best fits, we allow the parameters defining our broken power law Eddington ratio distribution to vary at each redshift, independent of each other. The variables defining the broken power law include the minimum Eddington ratio of the function ($\lambda_\text{min}$), the threshold between the two power laws ($\lambda_{\text{break}}$), and the slopes of the two power laws ($\alpha_1$, $\alpha_2$), as shown in Equation \ref{equ:bpl}. Our model was initially tested with a varying $\alpha_2$, however, we found little scatter in the resulting XLF fits and for simplicity have adopted the constant value of $\alpha_2=2.4$ to which our variable fits converged. This has the added benefit of limiting the number of free parameters. 

At each redshift we run the model multiple times with the following variations to the Eddington ratio distribution: $\alpha_1$ is allowed to vary between $0.0 \le \alpha_1 \le 1.0$ in steps of $0.01$, $\lambda_{\text{break}}$ is allowed to vary between $0.0 \le \log(\lambda_{\text{break}}) \le 4.0$ in steps of $0.1$, and $\lambda_\text{min}$ is allowed to vary between $-8.0 \le \log(\lambda_\text{min}) \le -4.5$ in steps of $0.5$. As such, there are only three parameters that may change at any time in order to reduce the number of possible degeneracies.
We examined the error introduced by fitting to a published XLF by running a Monte Carlo simulation of our model fit. Rather than fitting to the \citet{Air15} parametrization for a given redshift, we allow the binned estimates of the XLF for hard-X-rays to vary along their error bars. We determine the best-fit through $\chi^2$ minimization for each of these runs and find, e.g., the $1\sigma$ errors for the low-Eddington slope are $\delta \alpha_1\sim0.042$.
The results of these fits are shown in Figure \ref{fig:XLF} where we have selected a sample of redshifts covering the range studied by \citet{Jon17} and \citet{Air10}, allowing a comparison of our convolution (solid light colors) to the \citet{Air15} AGN XLF (solid black). For clarity, the \citet{Air15} $z=0$ line has been included at each redshift for comparison (dot dash grey).

Once we have determined the Eddington ratio distribution that produces the best-fit convolution, we use these parameters to define the AGN properties for each galaxy in the \citet{Mut13} model. This is very similar to what is done in \citet{Jon17}. 
Based on the best-fit Eddington ratio distribution at each redshift, we draw an intrinsic bolometric AGN luminosity.
We then apply the same bolometric correction to our bolometric AGN luminosities as in our convolution (\citealt{Don18}) to determine the AGN X-ray luminosity. 
With the ``best fit'' convolved model parameters we are able to simulate the luminosity distributions of our sample (dark colored dashed lines) and reproduce the \citet{Air15} AGN XLFs. 
Scatter at low redshifts at the highest luminosities is due to the relatively low number of extreme AGN in our volume limited sample. The X-ray emission from stellar processes and hot gas in our simulation is calculated following the prescription outlined in \citet{Jon17}, in which we utilize the \citet{Leh16} scaling relationships to determine the luminosity contribution from low mass X-ray binaries (LMXB) and high mass X-ray binaries (HMXB) from the \citet{Mut13} galaxy parameters. 
With knowledge of the dark matter, galaxy, and AGN properties of our intrinsic sample across the selected redshift range, we are able to compare the properties of the full AGN population to X-ray observations.

\section{Comparison of the Simulated AGN Population with X-ray Observations}\label{sec:xray}

\subsection{Investigating the co-evolution of galaxies and AGN}\label{ssec:evol}

\begin{figure}[!t]
\resizebox{80mm}{!}{\includegraphics{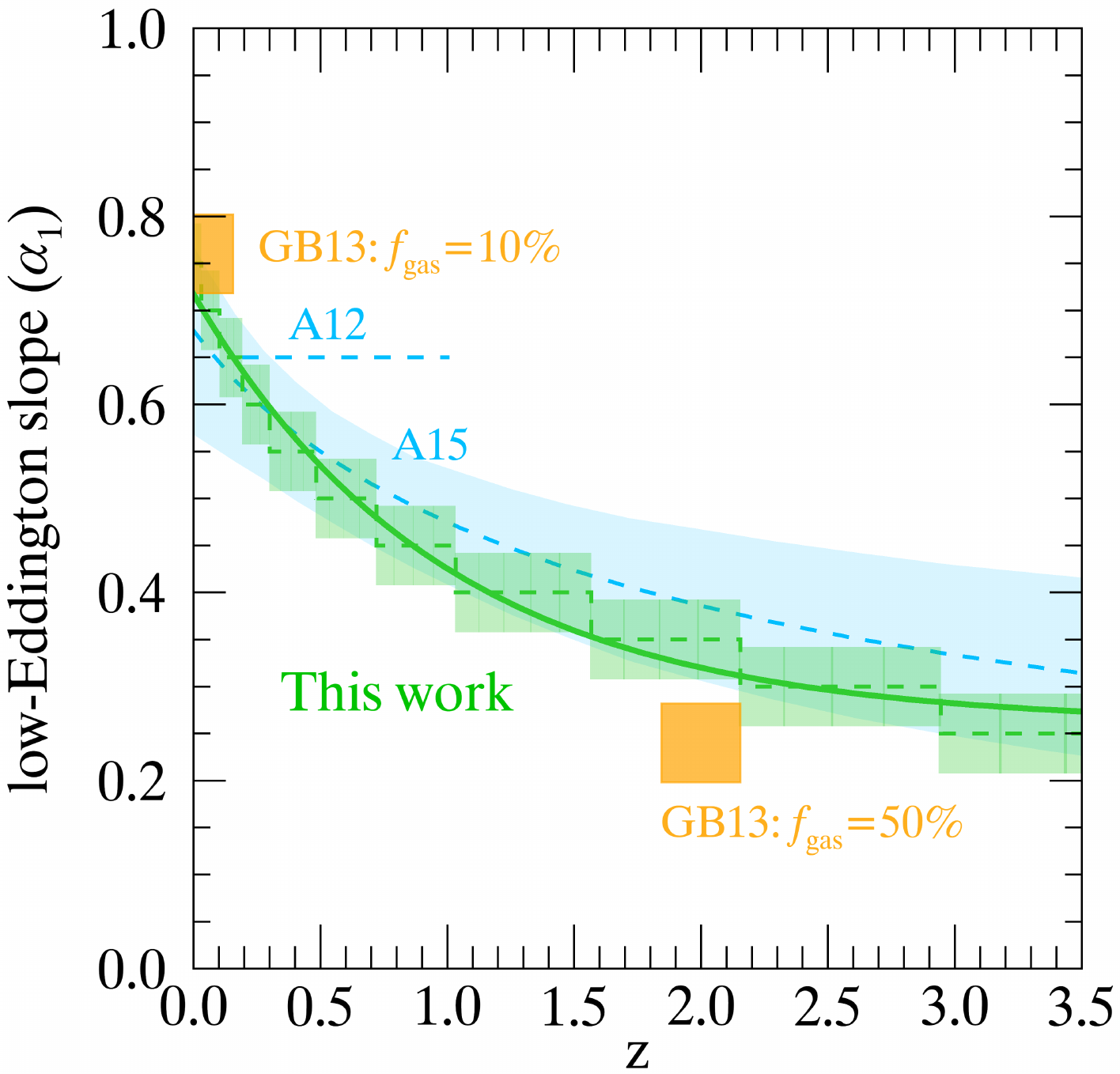}} \\
\caption{ Evolution of the Eddington ratio distribution slope as a function of redshift. We find that the low Eddington slope ($\alpha_1$) of our Eddington ratio distribution decreases quickly and flattens out as redshift increases and may be fit with an exponential function (green line; Equation \ref{equ:exp}). 
This may indicate a connection between the black hole activity and galaxy properties in which the accretion rate decreases as the gas fraction of the galaxy increases (\citealt{Gab13}; GB13). We include the GB13 simulated average Eddington ratio distribution slopes for gas-poor and gas-rich galaxies (orange regions) and find these to be qualitatively consistent with the evolution of our Eddington ratio distributions. We have also included the Aird et al. (2012; A12) power law slope ($0.2 < z < 1.0$) used by GB13, as well as the Aird et al. (2015; A15) Eddington ratio distribution for comparison (blue dashed lines). We expect our low-Eddington ratio distribution slopes to be consistent with those from A15 since our model is built to fit their observed XLFs.\label{fig:alpha}}
\end{figure}

\begin{figure}[!t]
\resizebox{85mm}{!}{\includegraphics{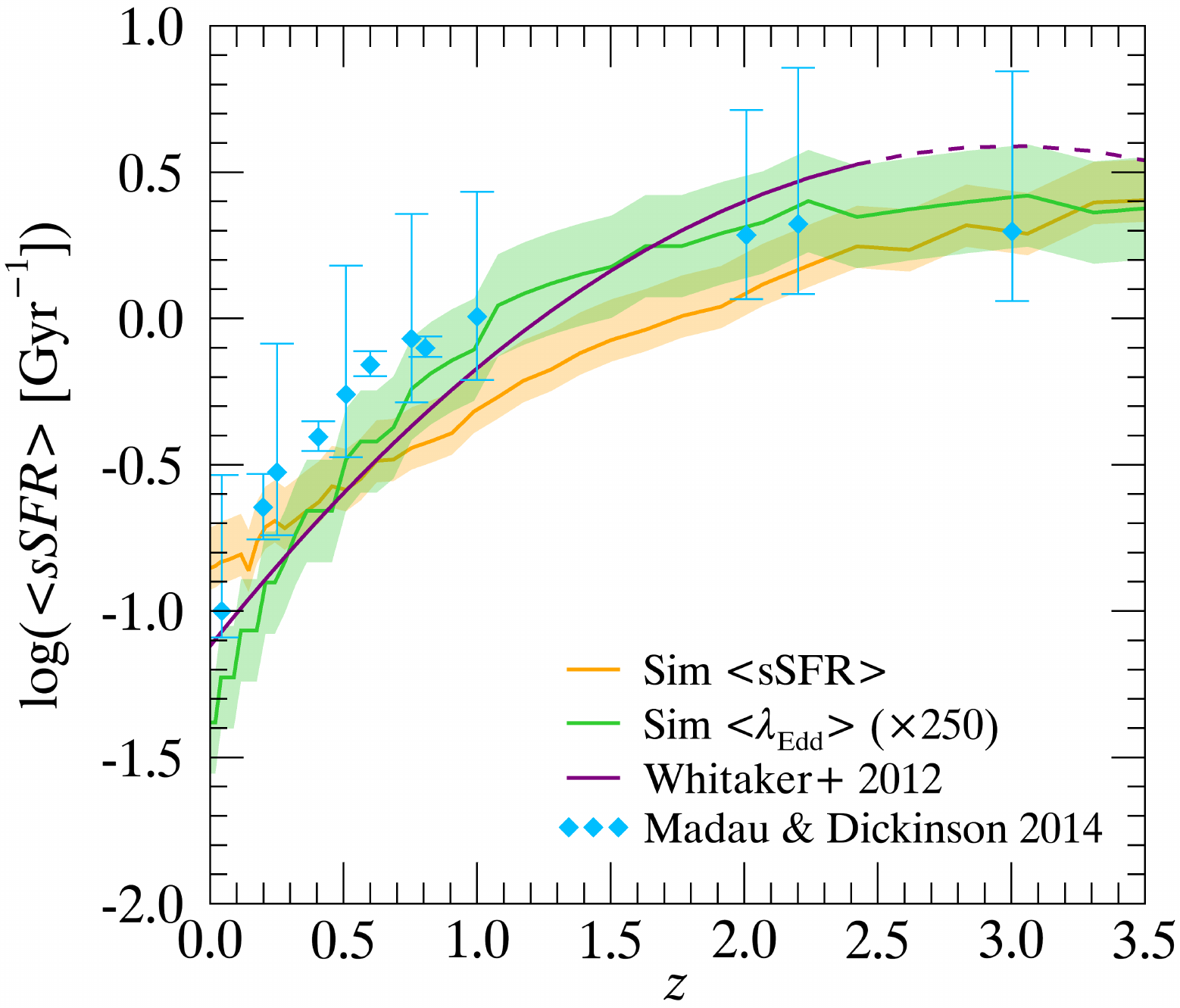}} \\
\caption{Evolution of the average specific star formation rate (sSFR) and Eddington ratio as a function of redshift. We find that the average Eddington ratio of our simulated sample increases quickly before beginning to flatten around a redshift of $z\sim1.5$ (green). This relationship is compared to the evolution of the sSFR in our simulation (orange). We find that the parameters exhibit similar evolution. We also compare these simulated evolutions to the parametrization of the star forming sequence from \citet{Whi12} (solid purple) and a compilation of sSFR from \citet{Mad14} and find a similar distribution. Note, the black hole accretion at the Eddington limit has been scaled by a factor of $250$ to directly compare to the sSFR histories. \label{fig:Edd}}
\end{figure}

The intrinsic black hole activity in our simulation provides insight into black hole-galaxy co-evolution as the best-fit parameters change across our redshift range. We focus on $\alpha_1$, the slope of the low-Eddington ratio distribution that directly determines the low end of the luminosity function through our convolution. Figure \ref{fig:alpha} shows how our $\alpha_1$ slope changes with redshift. 
We fit this relationship with the following exponential function:
\begin{equation}\label{equ:exp}
\alpha_1(z)=b_{0}~b_{1}^{~z}+b_{2}
\end{equation}
where the best-fit parameters are determined by the IDL routine \verb|COMFIT| to be: $b_{0}=0.49$, $b_{1}=0.41$, and $b_{2}=0.22$ (green line). This function may be used to further limit the number of free parameters defining the evolution of the low-Eddington slope. 
The flattening of the low-Eddington ratio distribution slope is not unexpected since the observed AGN X-ray luminosity functions are observed to flatten with increasing redshift. 
To fit this evolution, we convolve our black hole mass functions (which are defined by a constant stellar mass-black hole mass relationship) with an Eddington ratio distribution. In the case where the luminosity function slope is steeper at low luminosities compared to the low-mass regime of the black hole mass function, the fit is driven by the Eddington ratio distribution which will be steeper to compensate for the black hole mass function. While the low-luminosity regime of the luminosity function is not well constrained (e.g., \citealt{Air15,Buc15}), this pattern may point to a relationship between AGN activity and the evolving properties of the galaxies where AGN are found.

This relationship is clearer when put in the context of previous work (e.g., \citealt{Air12,Gab13}). \citet{Gab13} created a hydrodynamic simulation in which they investigate the Eddington ratio distribution for gas-poor (${f}_\text{gas}=10 \%$) and gas-rich (${f}_\text{gas}=50 \%$) galaxies, corresponding to typical star-forming galaxies at redshifts of $z=0$ and $z=2$, respectively (Figure \ref{fig:alpha}; orange regions). This distribution took the form of a power law, corresponding to the low-Eddington slope in our model, with $\alpha_1=0.76$ $(z=0)$ and $\alpha_1=0.24$ $(z=2)$ (following the nomenclature of Equation \ref{equ:bpl}). We find agreement with our evolving Eddington ratio distribution slopes and the \citet{Gab13} parameters based on a given galactic gas fraction. This is consistent with the \citet{Gab13} result that the Eddington ratio distribution slope decreases with increasing gas fraction (at higher redshifts). If we fit our simulation to an alternate XLF parametrization, such as \citet{Ued14}, we find that our low-Eddington ratio distribution slopes similarly decrease with increasing redshift out to $z~2$.


We can also investigate the evolution of the average Eddington ratio, or the black hole accretion history, for our simulated sample (Figure \ref{fig:Edd}; green). 
If black hole activity were tied to galactic activity, we expect to see a similar evolution in the specific star formation rates (sSFR; ratio of the star formation rate to the stellar mass) of our sample. It is possible to calculate the sSFR for our sample since the \citet{Mut13} model keeps track of the star formation rates for each individual galaxy. 
We probe this co-evolution by plotting the average sSFR of our galaxy sample (with stellar mass $\log{M}* > 9.0 {M}_\odot$) at each redshift, compared to the \citet{Whi12} parametrization of the star forming sequence, and a compilation of mean sSFR from \citet{Mad14}.
Directly comparing the average Eddington ratios to the global specific star formation rates typically requires scaling the average Eddington ratio by a constant value, in this case we use a scale factor of 250.

While the evolution of the growth of the black holes is similar in shape to the growth of the star forming global average, our black holes are growing slower than our galaxies. This is not surprising since our model assumes that all black holes are accreting with the same underlying Eddington ratio distribution and the most massive galaxies in our simulation are not star-forming. 
Differences in global and individual galaxy growth are also found when comparing the black hole-bulge relationship for individual galaxies to the global growth rate (similar growth, but the amplitudes are separated by $\sim10$ percent) (e.g., \citealt{Hec04,Kor13}).

We also find qualitative agreement between the shape of the average Eddington ratio evolution and our simulated average sSFR with the \citet{Whi12} line (Figure \ref{fig:Edd}; purple), and observed sSFR compiled by \citet{Mad14}. Recent work suggests there may be a difference in the integrated quantities of specific star formation rate and accretion (e.g., \citealt{Air15,Cap18,Yan18}). While we cannot compare directly to these works since we are calculating the average growth, there may be a similar discrepancy in the shapes of our specific star formation and accretion rate histories. However, within our error, we can only conservatively say that the average growth appears to be consistent.

\subsection{Host Galaxy SFR and Mass}\label{ssec:color}

\begin{figure}[!t]
\resizebox{85mm}{!}{\includegraphics{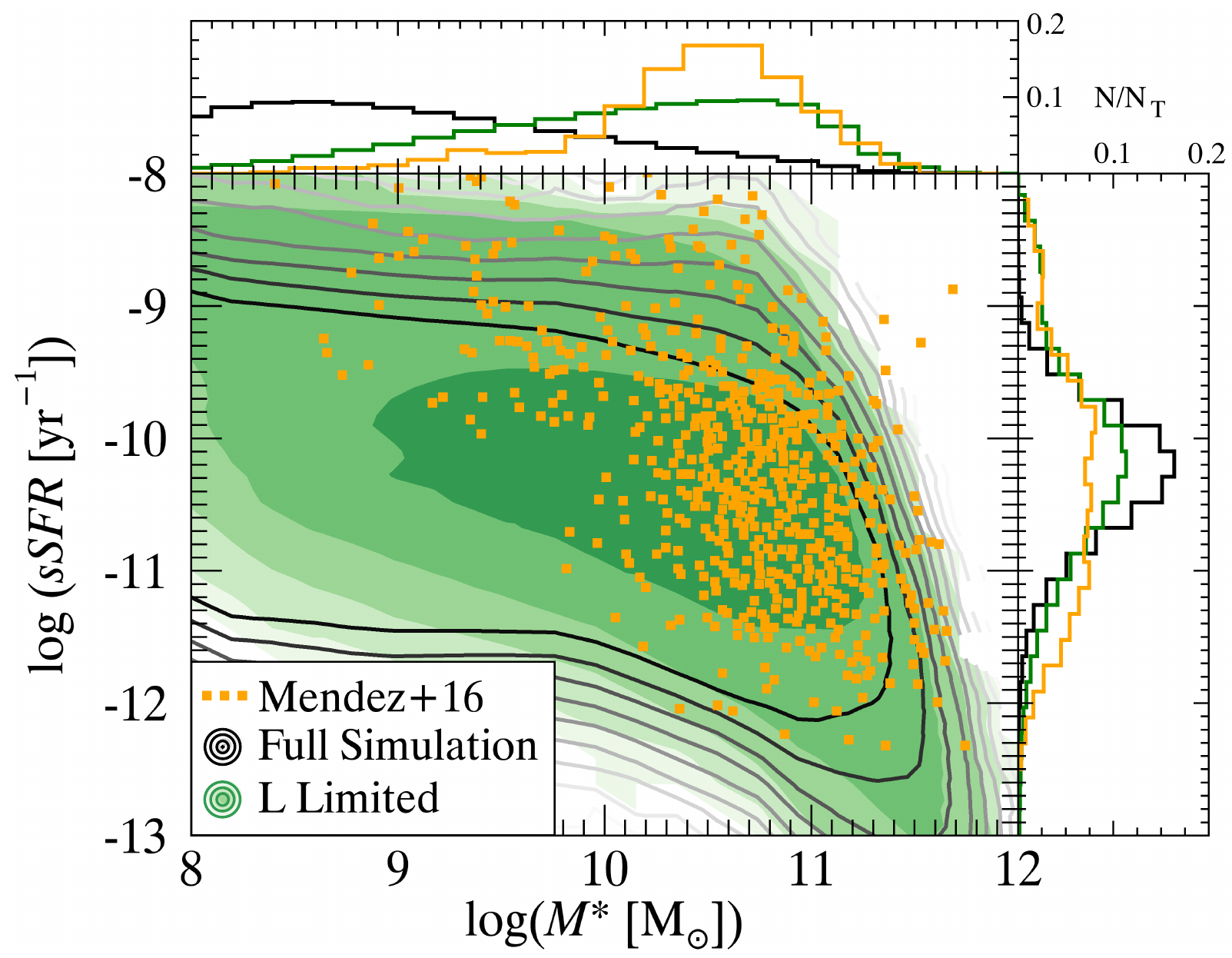}} \\
\caption{Specific star formation rate versus stellar mass for the full simulated sample of galaxies (black contours). We select our ``observed'' sample by introducing a luminosity threshold of ${L}_\text{X}=10^{41.5}$ (green contours) and compare this to the \citet{Men16} X-ray selected AGN with the same luminosity limit. We find the distributions of our simulated sample with added selection effects and the \citet{Men16} AGN cover the same dynamic range and are generally consistent at high stellar masses and across the wide range of sSFR. Normalized histograms for each sample are included on each axis for clarity. \label{fig:color}}
\end{figure}

In our model we simultaneously know the intrinsic properties of the black hole, galaxy, and dark matter halos. This allows us to investigate how observations of these properties are impacted by observational biases and selection effects. As in \citet{Jon17}, we examine the sSFR-M* distribution and compare it to surveys of observed X-ray selected AGN. Our intrinsic sSFR-M* distribution is shown in black contours in Figure \ref{fig:color}. 

\begin{figure}[!t]
\resizebox{85mm}{!}{\includegraphics{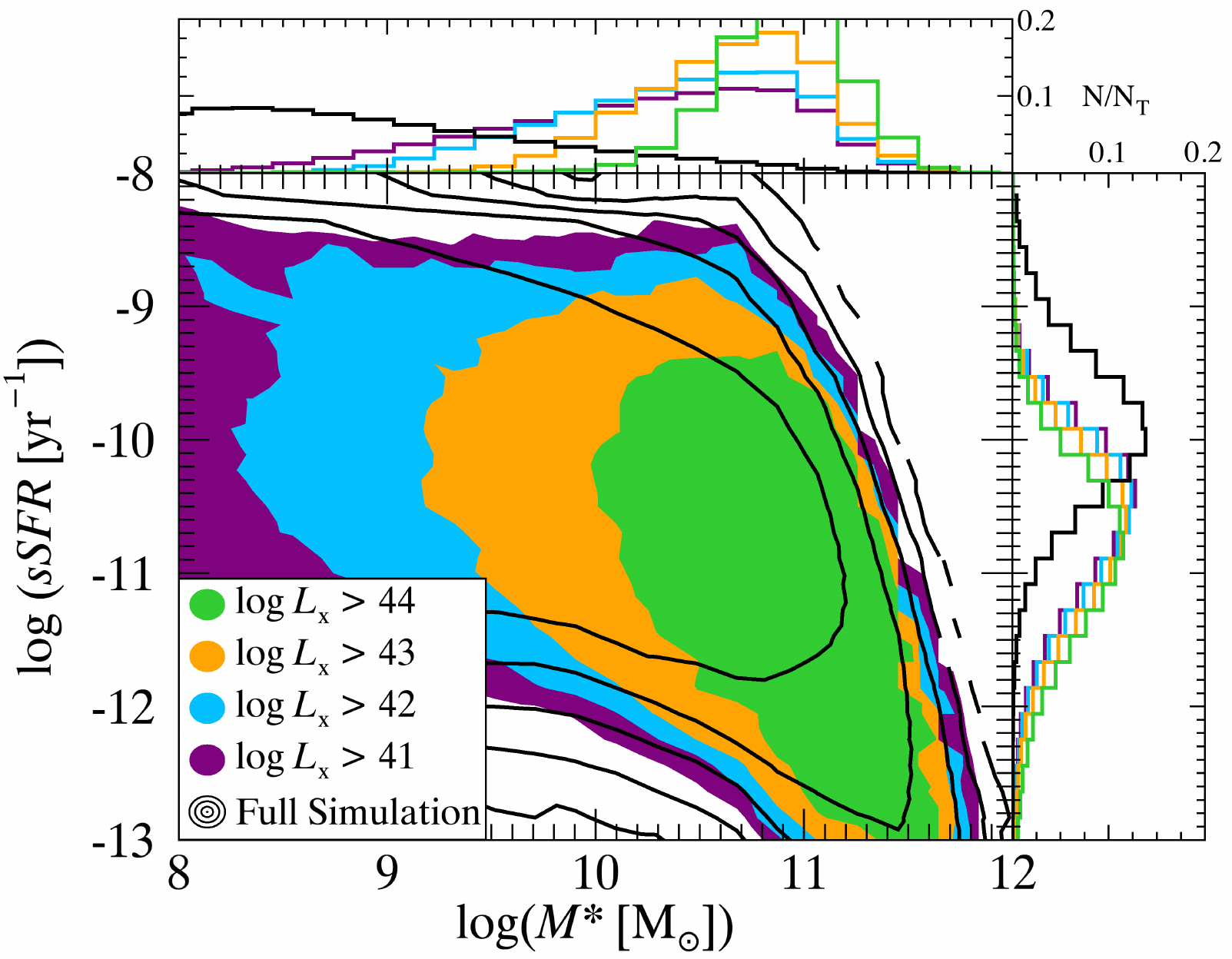}} \\
\caption{Illustration of the impact of AGN selection by luminosity limits on observed galaxy properties. Luminosity limits are shown in $\log({L}_\text{X})$. Raising the luminosity limit selects samples with increasingly higher stellar mass, as depicted by the contour boundaries for each given threshold. \label{fig:lim}}
\end{figure}

We again compare our simulation to the \citet{Men16} X-ray selected AGN (orange points). In this work, however, rather than applying a main sequence correction to the \citet{Men16} AGN and comparing it to our simulation at $z=0$, we select from our simulation the entire redshift range given by the \citet{Men16} X-ray selected AGN $0.2 < z < 1.2$. Once we select the snapshots corresponding to the survey redshifts, 
we add in obscuration following the same parametrization outlined in Jones et al. (2017) in which we assign obscuration based on the \citet{Mer14} incidence of obscuration and X-ray absorption corresponding to the NuSTAR-informed NH distribution (\citealt{Lan14,Lan15}). We then apply a luminosity threshold to ``select'' the AGN observed from our intrinsic sample.
The luminosity threshold we use mimics the \citet{Men16} limit of ${L}_\text{X}=10^{41.5}$. We find that the luminosity limited sample (green contours) covers the same dynamic range as the distribution of the Mendez X-ray selected AGN sample, more so than in \citet{Jon17} by including the survey redshift limits. Of particular note is the consistency at high stellar masses without the need for additional AGN suppression in high mass systems, keeping our model mass independent. We further illustrate the effects of luminosity limits on the observed galaxy properties in Figure \ref{fig:lim}, depicted by the contour boundaries for increasing luminosity thresholds. As the luminosity limit increases, the ``observed'' sample selected is ``biased'' towards higher stellar masses and correspondingly lower average sSFR (e.g., \citealt{Aza15,Air15}).

\subsection{Host Dark Matter Halos}\label{ssec:halo}

As with the host galaxies, this model allows us to probe the observed properties of dark matter halos for AGN selection techniques. This is a useful tool for investigating the impact of selection effects on clustering measurements as well as studies of large scale structure. We first compared our $z =0$ simulation to the observational clustering analysis of \citet{Ric13} and the simulation from \citet{Cha12}, as in \citet{Jon17}. While the results of this previous work were promising, this new analysis incorporates evolution and thus is better poised to directly compare to the dark matter halo distributions at the corresponding redshifts.

\begin{figure}[!t]
\resizebox{85mm}{!}{\includegraphics{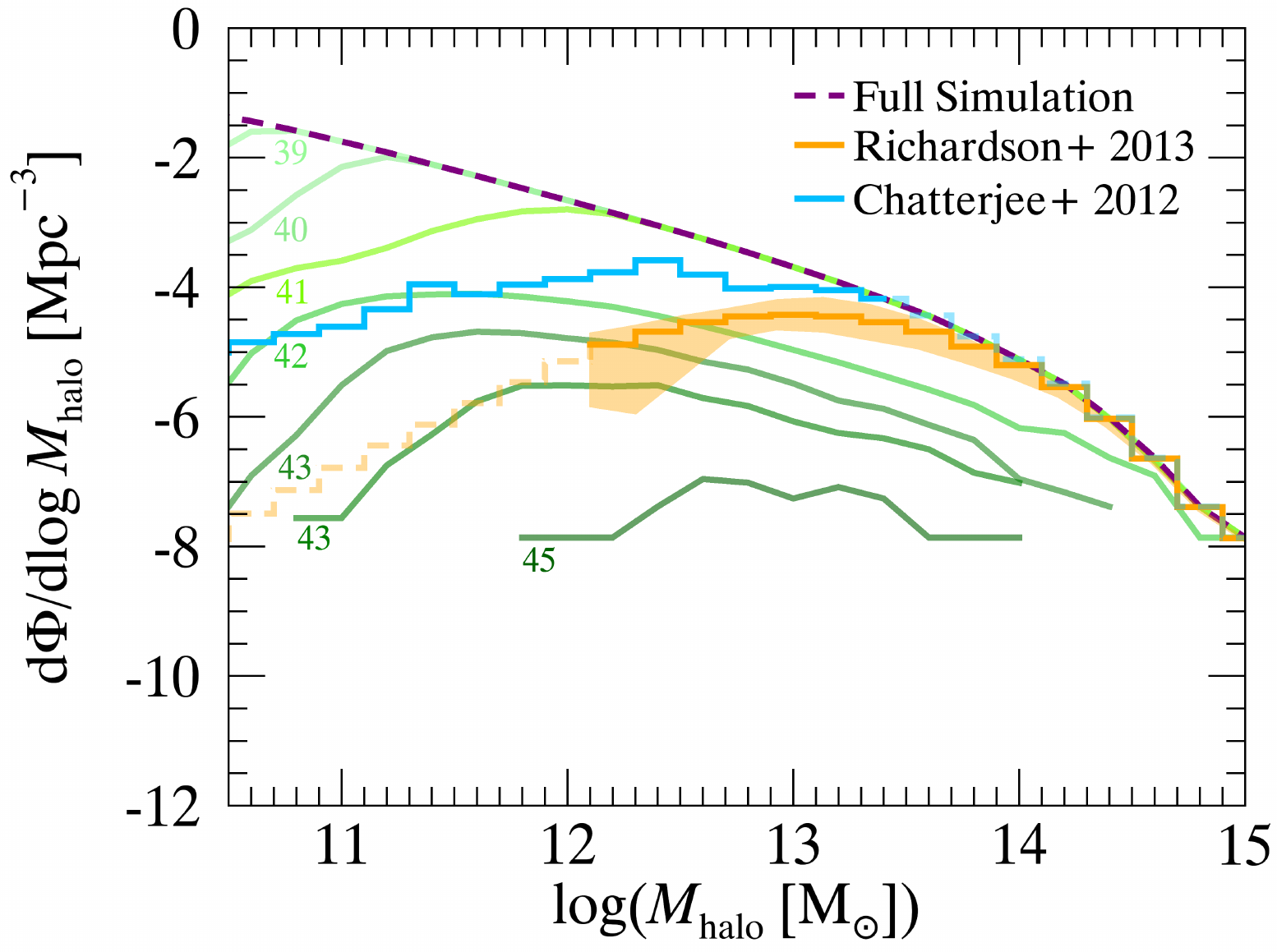}} \\
\caption{The dark matter halo mass distribution of the full simulated sample (dash purple) and luminosity limited samples ($39 \le \log{L}_\text{X} \ge 45$) of simulated AGN at $z = 1.0$ (green). We compare this mass distribution to the weighted halo occupation distributions (weighted by multiplying the HOD functions with the full population simulated halo mass distribution) of \citet{Cha12} and \citet{Ric13}, respectively. The luminosity limits of $\log({L}_\text{X}) > 41,42$ cover the same dynamic range as the \citet{Cha12} distribution (blue). Similarly we are able to recover parts of the \citet{Ric13} observational clustering analysis (orange). It is likely that the luminosity limits intrinsic to these observations are more complex than the constant limits shown.\label{fig:halo}}
\end{figure}

In this work, we directly compare the halo mass distributions of our simulated AGN rather than the HOD (the fraction of halos of a given mass that host an AGN). This takes into account the low density of halos at large masses by appropriately weighting the distribution at more common moderate halo masses. We multiply the halo mass function of our intrinsic sample at $z = 1.0$ with the HOD from \citet{Cha12} and \citet{Ric13} (Figure \ref{fig:halo}; blue, orange lines). This weighted distribution is compared with the mass distribution of the full simulated sample (dash purple). After applying obscuration to our simulated X-ray luminosities, we also select three luminosity limited samples ($39 \le \log{L}_\text{X} \ge 45$) at $z = 1.0$ (green) to compare to the theoretical and observationally motivated mass distributions. 
\begin{figure}[!t]
\resizebox{85mm}{!}{\includegraphics{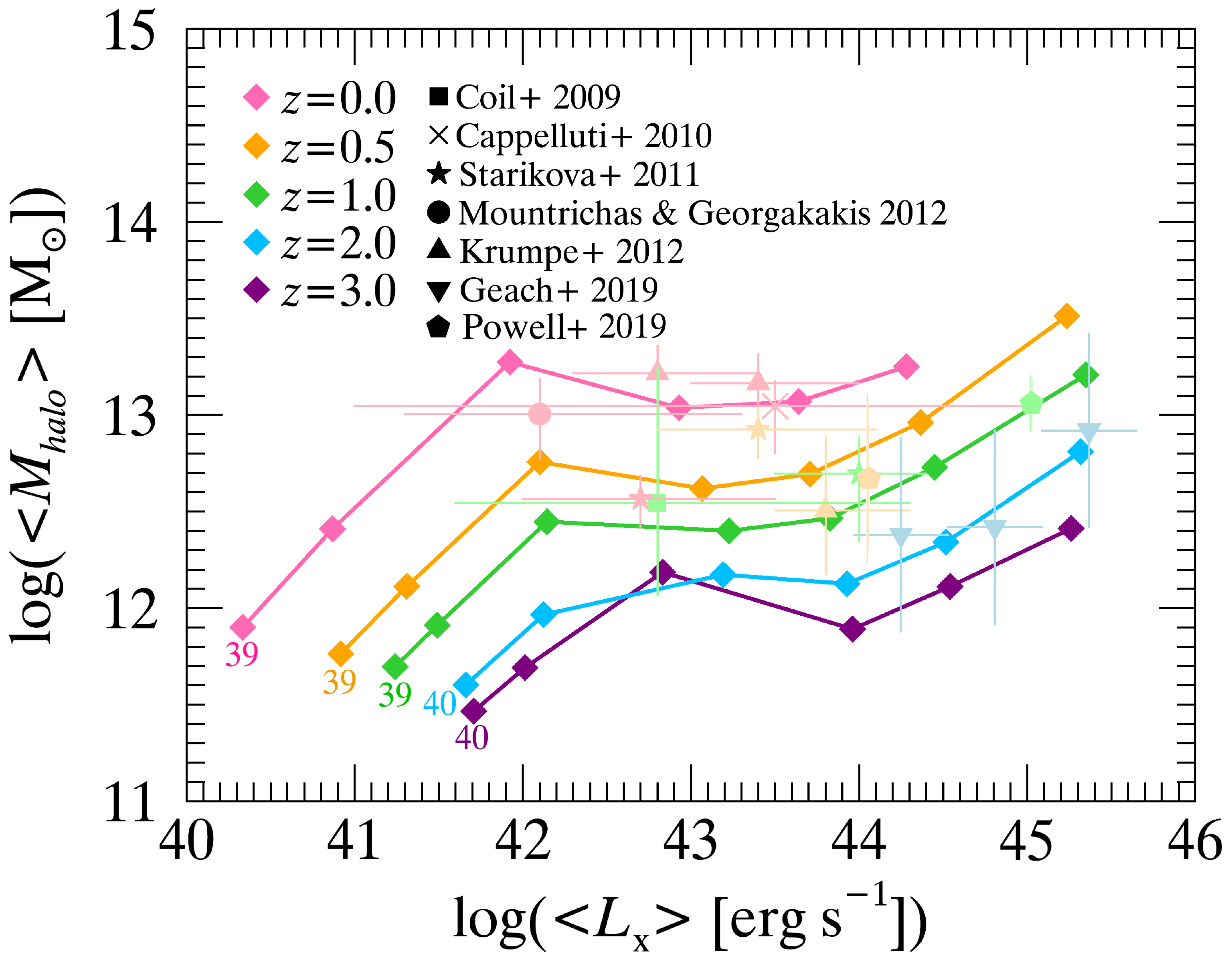}} \\
\caption{The mean dark matter halo mass versus the mean X-ray luminosity for a given X-ray luminosity limit at a selection of redshifts. The mean X-ray luminosity for each luminosity limit ($39 \le \log L_{\text{X}} \le 45$; steps of $1$ dex) is symbolized by diamonds with coordinating colors by redshift, while each observation is assigned its own symbol with a color that corresponds to the redshift color closest to its observed average redshift. We find that as the luminosity limit increases, the average simulated luminosity increases, while the average simulated halo mass experiences a plateau between $42 \le \log L_{\text{X}} < 44$ (corresponding to a luminosity limit between $41 \le \log L_{\text{X}} < 44$). This flat region is consistent with the observed halo properties of AGN in X-ray selected clustering surveys. \label{fig:MvL}}
\end{figure}

We find that as the luminosity limits increase, the dark matter halo mass distribution becomes increasingly ``biased'' toward higher halo masses. For a luminosity limit of $\log({L}_\text{X}) > 41.5$, we recover the \citet{Cha12} weighted halo mass distribution.
We are also able to recover parts of the observed \citet{Ric13} halo mass distribution (orange) using different luminosity limits. Our inability to recover the full halo mass distribution is not unexpected, since any biases that may impact this sample selection are likely more complicated than a single luminosity limit. 

We further investigate the impact of luminosity limits on our sample by calculating the average dark matter halo mass and luminosity as a function of luminosity limit and redshift. From the full simulation, five snapshots are selected that correspond to redshifts: $z\sim 0.0, 0.5, 1.0, 2.0, 3.0$.
At each redshift, we select a sub-sample based on luminosity limits between $39 \le \log L_{\text{X}} \le 45$ (in steps of 1 dex) and calculate the mean halo mass and X-ray luminosity.
We find, as expected, that as our luminosity limits increase the mean X-ray luminosity from our simulation also increases (diamonds; Figure \ref{fig:MvL}). Our mean halo masses, however, exhibit a plateau between $42 \le \log L_{\text{X}} \le 44$. This flattening is consistent with the average halo masses and luminosities of X-ray selected AGN in observational clustering surveys (e.g., \citealt{Coi09,Cap10,Sta11,Mou12,Kru12}, Powell et al. 2019 \textit{submitted}; select observed luminosities were renormalized to $L_{\text{X}}$[2-10 keV] in \citealt{Fan13cluster,Gea19}).
The observed plateau is a direct consequence of the Eddington limit: the only way to reach the highest luminosities is to have a large black hole accreting near its Eddington limit. Thus selecting AGN at high luminosities biases the sample to a narrower range of black hole masses and Eddington ratios. Likewise, at lower luminosities, there is a mix of high mass black holes accreting at low Eddington rates and low mass black holes accreting at high Eddington rates. As a result, there is a broader distribution in accretion rate and a weaker luminosity dependence for the average halo mass.

These results reaffirm the capability of our simple model to describe the properties of dark matter halos for luminosity limited AGN selection techniques. When comparing surveys and analyses of the environment where AGN reside, it is important to understand how the chosen selection methods impact the observed dark matter and host galaxy properties. This simulation provides insight into what subset of the AGN population is observed based on selection criteria.

\section{Conclusions}\label{sec:con}

The goal of this work is to investigate the evolution of AGN and their host galaxies using a simple prescription for AGN activity and a semi-analytic galaxy formation model from \citet{Mut13}. We have made improvements to the black hole accretion prescription from \citet{Jon17} by fitting the \citet{Air15} X-ray luminosity functions for redshifts $0.0 \le z < 4.12$. This involves convolving an intrinsic Eddington ratio distribution with the black hole mass functions determined directly from the galaxy formation histories, which decreases the computational cost of our model compared to that in \citet{Jon17}. From these best-fit Eddington ratio distributions we can build a simulated population of AGN corresponding to every galaxy in the \citet{Mut13} galaxy formation model. This allows us to simultaneously investigate the properties of the galaxy, AGN, and dark matter halos. 

The model described in this paper uses a forward-modeling technique in which we build a simulation of dark matter halos, galaxies, and AGN based on fiducial models of galaxy formation theory and AGN fueling. Our best-fit black hole accretion models are constrained by observations while our black hole mass functions are motivated by structure formation theory and the observed evolution of the galaxy stellar mass functions. With this prescription we are able to build a large volume of galaxy and AGN with knowledge of their properties as a whole, rather than the internal galaxy conditions. Additionally, we have selected a parametrization of the black hole accretion that is a universal shape (following the results of \citealt{Jon16}) which keeps our model simple and with few variables in order to reduce the number of degeneracies. This makes the simulation more straightforward to analyze and computationally inexpensive, running in sub-seconds, a significant improvement to more complicated semi-analytic models. 

Since we connect the dark matter halo mass to the galaxy stellar mass, the galaxy stellar mass to the black hole mass, and the black hole mass to the AGN activity, we can directly compare to observations and test for black hole-galaxy coevolution. A fundamental question that defines our model is \textit{what is the underlying relationship governing AGN activity required to recreate observations across cosmic time?} We use the black hole mass functions tied to the galaxy formation simulation convolved with a double power law to describe black hole accretion in order to model the observed \citet{Air15} XLF. This more directly tells us about the long term processes of black hole growth than observations because the volume of our simulation can be used to our advantage to smooth outliers by describing the typical black hole accretion distribution (\citealt{Hic14}). 

By calculating the intrinsic average AGN and galaxy activity in our evolving model, we can directly test relationships of black hole-galaxy co-evolution without the influence of observational biases. This may help dispel some of the uncertainty in the observational evidence of co-evolution due to current limitations observing the full AGN population (e.g., \citealt{Hic14,Pad17}). We find that the evolution of our black hole accretion distributions are consistent with the predicted evolution of galaxy gas fraction (\citealt{Gab13}), which is potential evidence for a black hole-galaxy connection through a common supply of gas. In addition, we find similar evolution of the average Eddington ratio distribution and the average specific star formation rate. This may be explained by a common supply of gas, or the presence of AGN and/or stellar feedback processes that may trigger or quench star formation and AGN activity. Furthermore, since our model also contains information about the dark matter halo and galaxy properties we can apply selection criteria to our full sample to compare directly to observations with similar limitations (e.g., volume, luminosity limit, mass limit) to investigate the impact of selection biases on the observed properties used in evidence of co-evolution.

The results of our analysis are summarized as follows:
\begin{itemize}
\item We find that the evolution of the slope defining the Eddington ratio distribution in our model becomes flatter with increasing redshift. This shape is consistent with the picture from \citet{Gab13} in which the black hole accretion rate varies based on the predicted behavior of galaxy gas fraction, where galaxies at higher redshifts have increased gas fraction and flatter Eddington ratio distributions. 

\item We find that the redshift evolution of the average Eddington ratio is broadly consistent with the evolution of the average specific star formation rate in galaxies, as well as the \citet{Whi12} star forming sequence. This co-evolution may be due to a common supply of gas, and/or feedback processes that may quench or trigger galaxy and black hole activity.

\item We confirm that selecting AGN based on a luminosity limit impacts the properties of the observed host galaxies and dark matter halos. We compare the distributions of the galaxy and halo properties of our luminosity limited simulated sample to observations of AGN selected samples (\citealt{Men16}) and dark matter halo clustering studies (\citealt{Cha12,Ric13}). We find that different AGN selection criteria yield different galaxy and halo properties.
\end{itemize}

We have shown that a simple model of black hole accretion and galaxy evolution is able to reproduce the observed evolution of AGN and black holes across cosmic time. Given the breadth of the available AGN, galaxy, and dark matter observables built into this simulation, the natural next step would be to investigate additional properties and selection effects on the X-ray AGN population. 
This work has further confirmed the universality of a simple accretion model to describe black hole growth, but has also shown there can be some evolution in those model parameters (e.g., with redshift) that may be connected to host galaxy properties. Based on the assumption of a universal Eddington ratio distribution to describe the full multi-wavelength AGN population, it is possible to expand this work into a powerful multi-wavelength simulation to make predictions for the next generation of observatories.

\acknowledgments
We thank our collaborators for their insights in developing this work, as well as for constructive comments that improved the paper. We also thank our referee for comments that improved this work.
This work was supported in part by the National Aeronautics and Space Administration under Grant Number NNX15AU32H issued through the NASA Education Minority University Research Education Project (MUREP) through the NASA Harriett G. Jenkins Graduate Fellowship activity. This research has made use of NASA's Astrophysics Data System.


\bibliography{ap_evol}


\end{document}